**Matrix genetics, part 3: the evolution of the genetic code from the viewpoint of the genetic octave Yin-Yang-algebra**

Sergey V. Petoukhov
Department of Biomechanics, Mechanical Engineering Research Institute of the Russian Academy of Sciences
petoukhov@hotmail.com, petoukhov@imash.ru, http://symmetry.hu/isabm/petoukhov.html

**Corresponding address**: Kutuzovskiy prospect, d. 1/7, kv.58, Moscow, 121248, Russia.

**Abstract**. The set of known dialects of the genetic code (GC) is analyzed from the viewpoint of the genetic octave Yin-Yang-algebra. This algebra was described in the previous author's publications. The algebra was discovered on the basis of structural features of the GC in the matrix form of its presentation ("matrix genetics"). The octave Yin-Yang-algebra is considered as the pre-code or as the model of the GC. From the viewpoint of this algebraic model, for example, the sets of 20 amino acids and of 64 triplets consist of sub-sets of "male", "female" and "androgynous" molecules, etc. This algebra permits to reveal hidden peculiarities of the structure and evolution of the GC and to propose the conception of "sexual" relationships among genetic molecules. The first results of the analysis of the GC systems from such algebraic viewpoint say about the close connection between evolution of the GC and this algebra. They include 8 evolutionary rules of the dialects of the GC. The evolution of the GC is appeared as the struggle between male and female beginnings. The hypothesis about new biophysical factor of "sexual" interactions among genetic molecules is put forward. The matrix forms of presentation of elements of the genetic octave Yin-Yang-algebra are connected with Hadamard matrices by means of the simple U-algorithm. Hadamard matrices play a significant role in the theory of quantum computers, in particular. It gives new opportunities for possible understanding the GC systems as quantum computer systems. Revealed algebraic properties of the GC permit to put forward the problem of algebraization of bioinformatics on the basis of the algebras of the GC. Our investigation is connected with the question: what is life from the viewpoint of algebra? The algebraic version of the origin of the GC is discussed.

## 1 Introduction

This article is devoted to the first results of investigations of the evolution of the genetic code (GC) from the viewpoint of the genetic octave Yin-Yang-algebra, which was described in the article [Petoukhov, arXiv:0803.3330] and in the book [Petoukhov, 2008]. This genetic algebra defines the system of 8-dimensional numbers $YY_8$ (the matrix form of presentation of these numbers $YY_8$ is presented on Figures 1, 2):

$$YY_8 = x_0*\mathbf{f_0}+x_1*\mathbf{m_1}+x_2*\mathbf{f_2}+x_3*\mathbf{m_3}+x_4*\mathbf{f_4}+x_5*\mathbf{m_5}+x_6*\mathbf{f_6}+x_7*\mathbf{m_7} \quad (1)$$

Multiplication of any two members of such octave numbers $YY_8$ generates a new octave number of the same system. This situation is similar to the situation of real numbers (or of complex numbers, or of hypercomplex numbers) when multiplication of any two members of the numeric system generates a new member of the same numerical system. In other words, we have received the new numerical system of octave numbers $YY_8$ from the natural structure of the genetic code.

This numeric system has regular and sharp distinctions between the sub-set of the basic elements $\mathbf{f_0}$, $\mathbf{f_2}$, $\mathbf{f_4}$, $\mathbf{f_6}$ and the sub-set of the basic elements $\mathbf{m_1}$, $\mathbf{m_3}$, $\mathbf{m_5}$, $\mathbf{m_7}$. These distinctions are based on attributes of the squares of these elements (see the multiplication table on Figure 2). The squares of the $\mathbf{f_0}$, $\mathbf{f_2}$, $\mathbf{f_4}$, $\mathbf{f_6}$ are equal to $\pm\mathbf{f_0}$ always. The basic element $\mathbf{f_0}$ possesses all properties of the real unit in relation to each member of the first sub-set of the basic elements $\mathbf{f_0}$, $\mathbf{f_2}$, $\mathbf{f_4}$, $\mathbf{f_6}$, but it does not possess such properties in relation to the members of the second sub-set $\mathbf{m_1}$, $\mathbf{m_3}$, $\mathbf{m_5}$, $\mathbf{m_7}$. In this reason the element $\mathbf{f_0}$ is named the quasi-real unit from the first sub-set of the basic elements. The squares of the members of the second sub-set $\mathbf{m_1}$, $\mathbf{m_3}$, $\mathbf{m_5}$, $\mathbf{m_7}$ are equal to $\pm\mathbf{m_1}$. The basic element $\mathbf{m_1}$ possesses all properties of the real unit in relation to each member of the second sub-set $\mathbf{m_1}$, $\mathbf{m_3}$, $\mathbf{m_5}$, $\mathbf{m_7}$, but it does not possess such properties in relation to the members of the first sub-set $\mathbf{f_0}$, $\mathbf{f_2}$, $\mathbf{f_4}$, $\mathbf{f_6}$. In this reason the element $\mathbf{m_1}$ is named the quasi-real unit from the second sub-set of the basic elements.

It was unexpected that all members of the first (second) sub-set of the basic elements $\mathbf{f_0}$, $\mathbf{f_2}$, $\mathbf{f_4}$, $\mathbf{f_6}$ ($\mathbf{m_1}$, $\mathbf{m_3}$, $\mathbf{m_5}$, $\mathbf{m_7}$) together with their coordinates $x_0$, $x_2$, $x_4$, $x_6$ ($x_1$, $x_3$, $x_5$, $x_7$) have the even (odd) indexes as the formal attribute for their comfortable identification. Moreover the elements and coordinates from the first sub-set $\mathbf{f_0}$, $\mathbf{f_2}$, $\mathbf{f_4}$, $\mathbf{f_6}$ are disposed in the columns with the even numbers 0, 2, 4, 6 in the table of multiplication of the basic elements and in the matrix form of presentation of the numbers $YY_8$ (see Figure 2). The elements and coordinates from the second sub-set $\mathbf{m_1}$, $\mathbf{m_3}$, $\mathbf{m_5}$, $\mathbf{m_7}$ are disposed in the columns with the odd numbers 1, 3, 5, 7.

| | | | | | | | |
|---|---|---|---|---|---|---|---|
| **CCC**<br>**Pro, $x_0$** | **CCA**<br>**Pro, $x_1$** | **CAC**<br>**His, $-x_2$** | **CAA**<br>**Gln, $-x_3$** | **ACC**<br>**Thr, $x_4$** | **ACA**<br>**Thr, $x_5$** | **AAC**<br>**Asn, $-x_6$** | **AAA**<br>**Lys, $-x_7$** |
| **CCU**<br>**Pro, $x_0$** | **CCG**<br>**Pro, $x_1$** | **CAU**<br>**His, $-x_2$** | **CAG**<br>**Gln, $-x_3$** | **ACU**<br>**Thr, $x_4$** | **ACG**<br>**Thr, $x_5$** | **AAU**<br>**Asn, $-x_6$** | **AAG**<br>**Lys, $-x_7$** |
| **CUC**<br>**Leu, $x_2$** | **CUA**<br>**Leu, $x_3$** | **CGC**<br>**Arg, $x_0$** | **CGA**<br>**Arg, $x_1$** | **AUC**<br>**Ile, $-x_6$** | **AUA**<br>**Met, $-x_7$** | **AGC**<br>**Ser, $-x_4$** | **AGA**<br>**Stop, $-x_5$** |
| **CUU**<br>**Leu, $x_2$** | **CUG**<br>**Leu, $x_3$** | **CGU**<br>**Arg, $x_0$** | **CGG**<br>**Arg, $x_1$** | **AUU**<br>**Ile, $-x_6$** | **AUG**<br>**Met, $-x_7$** | **AGU**<br>**Ser, $-x_4$** | **AGG**<br>**Stop, $-x_5$** |
| **UCC**<br>**Ser, $x_4$** | **UCA**<br>**Ser, $x_5$** | **UAC**<br>**Tyr, $-x_6$** | **UAA**<br>**Stop, $-x_7$** | **GCC**<br>**Ala, $x_0$** | **GCA**<br>**Ala, $x_1$** | **GAC**<br>**Asp, $-x_2$** | **GAA**<br>**Glu, $-x_3$** |
| **UCU**<br>**Ser, $x_4$** | **UCG**<br>**Ser, $x_5$** | **UAU**<br>**Tyr, $-x_6$** | **UAG**<br>**Stop, $-x_7$** | **GCU**<br>**Ala, $x_0$** | **GCG**<br>**Ala , $x_1$** | **GAU**<br>**Asp, $-x_2$** | **GAG**<br>**Glu, $-x_3$** |
| **UUC**<br>**Phe, $-x_6$** | **UUA**<br>**Leu, $-x_7$** | **UGC**<br>**Cys, $-x_4$** | **UGA**<br>**Trp, $-x_5$** | **GUC**<br>**Val, $x_2$** | **GUA**<br>**Val, $x_3$** | **GGC**<br>**Gly, $x_0$** | **GGA**<br>**Gly, $x_1$** |
| **UUU**<br>**Phe, $-x_6$** | **UUG**<br>**Leu, $-x_7$** | **UGU**<br>**Cys, $-x_4$** | **UGG**<br>**Trp, $-x_5$** | **GUU**<br>**Val, $x_2$** | **GUG**<br>**Val, $x_3$** | **GGU**<br>**Gly, $x_0$** | **GGG**<br>**Gly, $x_1$** |

Figure 1. The genetic matrix $P^{(3)}=[C\ A;\ U\ G]^{(3)}$ of 64 triplets, 20 amino acids and 4 stop-codons for the vertebrate mitochondria genetic code. The black-and-white mosaic of the matrix reflects the degeneracy of the code (see [Petoukhov, arXiv:0802.3366]). The Yin-coordinates $x_0$, $x_2$, $x_4$, $x_6$ and the Yang-coordinates $x_1$, $x_3$, $x_5$, $x_7$ of the octave Yin-Yang-algebra are shown additionally (from [Petoukhov, arXiv:0803.3330]). The black cells contain the coordinates with the sign „+" and the white cells contain the coordinates with the sign „-".

The Pythagorean and Ancient Chinese traditions considered even numbers as Yin-numbers (or female numbers) and odd numbers as Yang-numbers (or male numbers). In this reason this genetic algebra was named conditionally as the octave Yin-Yang-algebra (or the even-odd-algebra or the bisex-algebra). Correspondingly the elements with the even indexes 0, 2, 4, 6 are named the female elements (or the Yin-elements) and the elements with the odd indexes 1, 3, 5, 7 are named the male elements (or the Yang-elements).

Figure 1 shows the genetic matrix $P^{(3)}=[C\ A;\ U\ G]^{(3)}$ of 64 triplets with their code meanings for the case of the vertebrate mitochondria genetic code, which is considered as the basic dialect of the genetic code. This genetic matrix (or "genomatrix" briefly) possesses the black-and-white mosaic, which reflects the specifics of the degeneracy of the code and which has special symmetric properties. This genetic matrix is transformed (by means of the "alphabetic" algorithm of the Yin-Yang-digitization of 64 triplets) into the matrix form of presentation of elements $YY_8$ (1) of the genetic octave Yin-Yang-algebra (see details in [Petoukhov, arXiv:0803.3330]).

If the matrix on Figure 1 has the eight real coordinates $x_0$, $x_1$, …, $x_7$ in its cells only, we have the matrix $YY_8$ (Figure 2, on the right side) as the matrix form of presentation of new numeric system of 8-dimensional numbers $YY_8$ (1), which can be named genetic octaves ([Petoukhov, arXiv:0803.3330]). Figure 2 shows the multiplication table of the basic elements of this numeric system (or its algebra over field).

| | №0 | №1 | №2 | №3 | №4 | №5 | №6 | №7 |
|---|---|---|---|---|---|---|---|---|
| | $f_0$ | $m_1$ | $f_2$ | $m_3$ | $f_4$ | $m_5$ | $f_6$ | $m_7$ |
| $f_0$ | $f_0$ | $m_1$ | $f_2$ | $m_3$ | $f_4$ | $m_5$ | $f_6$ | $m_7$ |
| $m_1$ | $f_0$ | $m_1$ | $f_2$ | m3 | $f_4$ | $m_5$ | $f_6$ | $m_7$ |
| $f_2$ | $f_2$ | $m_3$ | $-f_0$ | $-m_1$ | $-f_6$ | $-m_7$ | $f_4$ | $m_5$ |
| $m_3$ | $f_2$ | $m_3$ | $-f_0$ | $-m_1$ | $-f_6$ | $-m_7$ | $f_4$ | $m_5$ |
| $f_4$ | $f_4$ | $m_5$ | $f_6$ | $m_7$ | $f_0$ | $m_1$ | $f_2$ | $m_3$ |
| $m_5$ | $f_4$ | $m_5$ | $f_6$ | $m_7$ | $f_0$ | $m_1$ | $f_2$ | $m_3$ |
| $f_6$ | $f_6$ | $m_7$ | $-f_4$ | $-m_5$ | $-f_2$ | $-m_3$ | $f_0$ | $m_1$ |
| $m_7$ | $f_6$ | $m_7$ | $-f_4$ | $-m_5$ | $-f_2$ | $-m_3$ | $f_0$ | $m_1$ |

| №0 | №1 | №2 | №3 | №4 | №5 | №6 | №7 |
|---|---|---|---|---|---|---|---|
| $x_0$ | $x_1$ | $-x_2$ | $-x_3$ | $x_4$ | $x_5$ | $-x_6$ | $-x_7$ |
| $x_0$ | $x_1$ | $-x_2$ | $-x_3$ | $x_4$ | $x_5$ | $-x_6$ | $-x_7$ |
| $x_2$ | $x_3$ | $x_0$ | $x_1$ | $-x_6$ | $-x_7$ | $-x_4$ | $-x_5$ |
| $x_2$ | $x_3$ | $x_0$ | $x_1$ | $-x_6$ | $-x_7$ | $-x_4$ | $-x_5$ |
| $x_4$ | $x_5$ | $-x_6$ | $-x_7$ | $x_0$ | $x_1$ | $-x_2$ | $-x_3$ |
| $x_4$ | $x_5$ | $-x_6$ | $-x_7$ | $x_0$ | $x_1$ | $-x_2$ | $-x_3$ |
| $-x_6$ | $-x_7$ | $-x_4$ | $-x_5$ | $x_2$ | $x_3$ | $x_0$ | $x_1$ |
| $-x_6$ | $-x_7$ | $-x_4$ | $-x_5$ | $x_2$ | $x_3$ | $x_0$ | $x_1$ |

Figure 2. On the left: the table of multiplication of the basic elements $f_0$, $m_1$, $f_2$, $m_3$, $f_4$, $m_5$, $f_6$, $m_7$ of the octave numbers $YY_8$ (1). On the right: the matrix form of presentation of the numbers $YY_8$ (the brief version of the matrix on Figure 1).

This genetic octave Yin-Yang-algebra is penetrated by the principle of binary opposition of elements with even and odd indexes. But one can note that the principle of binary opposition penetrates many systems of the genetic code. Really, DNA has the double spiral configuration; each letter of genetic alphabet has its binary-oppositional partner in a complimentary pair; amino acids have amphoteric properties (they demonstrate acid properties and alkaline properties simultaneously; a non-dissociated form of amino acids is transformed in a dipolar form under conditions of neutral water solution); etc. It seems that many of such facts of binary oppositions in genetic systems possess hidden connections with the genetic Yin-Yang-algebra which exists not accidentally.

Figure 3 shows initial analogies and interrelations between the genomatrix $P^{(3)}=[C\ A;\ U\ G]^{(3)}$ and the matrix $YY_8$ (Figures 1, 2). These data testify into favor that this octave Yin-Yang-algebra is the adequate model of the genetic code. Each triplet, which is disposed together with one of the female YY-coordinates $x_0$, $x_2$, $x_4$, $x_6$ in a mutual matrix cell (Figure 1), is named the female triplet or the Yin-triplet. The third position of all female triplets is occupied by the pyrimidine C or U/T. In this reason the female triplets can be named "pyrimidine triplets" as well. Each triplet, which is disposed together with one of the male YY-coordinates $x_1$, $x_3$, $x_5$, $x_7$ in a mutual matrix cell, is named the male triplet or the Yang-triplet. The third position of all male triplets is occupied by the purine A or G. In this reason the male triplets can be named "purine triplets". In such algebraic way the whole set of 64 triplets was divided into two sub-sets of Yin-triplets and Yang-triplets (see details in [Petoukhov, arXiv:0803.3330]).

| **The octet genomatrix $P^{(3)}$=[C A; U G]$^{(3)}$** | **The Yin-Yang octet matrix $YY_8$** |
|---|---|
| This genomatrix has the binary mosaic of symmetrical character with 32 black triplets and with 32 white triplets. | This matrix has the same mosaic, which contains 32 YY-coordinates with the sign "+" and 32 YY-coordinates with the sign "-" |
| The pair of neighboring rows under numbers 0-1, 2-3, 4-5, 6-7 are equivalent to each other by a disposition of identical amino acids. | The pair of neighboring rows under numbers 0-1, 2-3, 4-5, 6-7 are equivalent to each other by a disposition of identical YY-coordinates. |
| The pyrimidine triplets occupy the columns with even numbers 0, 2, 4, 6. The purine triplets occupy the columns with the odd numbers 1, 3, 5, 7. | The YY-coordinates $x_0$, $x_2$, $x_4$, $x_6$, which are connected with the sub-set of the basic elements with the quasi-real unit $f_0$, occupy the columns with even numbers 0, 2, 4, 6. The YY-coordinates $x_1$, $x_3$, $x_5$, $x_7$, which are connected with the sub-set of the basic elements with the quasi-real unit $m_1$, occupy the columns with the odd numbers 1, 3, 5, 7. |
| The half of kinds of amino acids is presented in the quadrants along the main matrix diagonal only (Ala, Arg, Asp, Gln, Glu, Gly, His, Leu, Pro, Val). The second half of kinds of amino acids is presented in the quadrants along the second diagonal only (Asn, Cys, Ile, Lys, Met, Phe, Ser,Thr, Trp, Tyr). | The half of kinds of YY-coordinates ($x_0$, $x_1$, $x_2$, $x_3$) is presented in the quadrants along the main matrix diagonal only. The second half of kinds of YY-coordinates ($x_4$, $x_5$, $x_6$, $x_7$) is presented in the quadrants along the second coordinates only. |
| If each of the black matrix cells has the number "+1" and each of white matrix cells has the number "-1", then the multiplication of the $P^{(3)}$ with itself produces the tetra-self-reproduction of this matrix $\{[C\ A;\ U\ G]^{(3)}\}^2=4*[C\ A;\ U\ G]^{(3)}$, which reminds of the tetra-reproduction of the gametal cells in meiosis. [Petoukhov, 2008b; arXiv:0803.0888, arXiv:0803.3330] | If each of the YY-coordinates is equal to 1 ($x_0=x_1=\ldots=x_7=1$), then the multiplication of the octave number $YY_8$ (1) with itself produces its quadrupling $YY_8^2=4*YY_8$, which corresponds to the tetra-self-reproduction of the matrix presentation of this number [Petoukhov, 2008b; arXiv:0803.3330. |
| All 6 variants of dispositions of triplets, which are produced by permutations of the three positions inside triplets, produce 6 individual (8x8)-matrices with new symmetrical dispositions of black triplets and white triplets together with appropriate high-degeneracy amino acids, low-degeneracy amino acids and stop-signals there [Petoukhov, 2008b; arXiv:0803.0888, arXiv:0803.3330]. | All 6 variants of matrix dispositions of YY-coordinates $x_0$, $x_1$, …,$x_7$ , which correspond to the 6 variants of the dispositions of the appropriate triplets with permutations of its three positions, produce 6 octave Yin-Yang-numeric systems with their matrix presentations in a form of the appropriate individual (8x8)-matrices [Petoukhov, 2008b; arXiv:0803.3330]. |

Figure 3. Examples of the conformity between the genomatrix [C A; U G]$^{(3)}$ and the matrix $YY_8$.

The data of the table on Figure 3 do not exhaust the interconnections between the genetic code systems and the Yin-Yang matrices [Petoukhov, 2008b]. This matrix form of presentation of the genetic Yin-Yang-algebra (Figures 1, 2) was received in the course of author's attempts to understand a possible basis of the symmetric black-and-white mosaic, which reflects the specifics of the degeneracy of the genetic code. In the result of these attempts such algebraic model of the genetic code was received, the matrix of which coincides with the initial genetic matrix $P^{(3)} = [C\ A;\ U\ G]^{(3)}$ not only in the black-and-white mosaic, but in some other significant features also.

## 2 The evolution of the genetic code from the viewpoint of the genetic octave Yin-Yang-algebra

The theme of male and female beginnings and biological reproduction connected with them is one of the main themes in human civilization. This binary opposition exists in different forms in many theories in the fields of psychology, biology, culture, religions, etc. (male and female types in psychology; male and female chromosomes; male and female gametal cells; etc.). It is considered ordinary that male and female beginnings in the nature are necessary to continue life and its development [Bull,1983; Geodakian, 1978, 1999; Karlin, Lessard, 1986; Maynard Smith, 1978; Mooney, 1992; Williams, 1975]. The tendency of thinkers to reflect the natural fact of male and female beginnings on a formal language is known from the ancient time. For example, thoughts about fundamental meanings of male and female beginnings are reflected by thinkers of Ancient China and of the Pythagorean School into the thematic division of the series of natural numbers, where even numbers embody the female beginning (Yin) and odd numbers embody the male beginning (Yang).

Inheritance of sexual attributes exists in living nature. The results of investigations in the field of matrix genetics permits to suppose the following: 1) ensembles of binary-oppositional attributes, which exist in molecular-genetic systems, are related with the problem of male and female beginnings; 2) many phenomenological features of these ensembles can be expressed on the language of multidimensional numbers, first of all, on the language of the genetic octave Yin-Yang-algebra which gives new possibilities to investigate such binary oppositions.

Taking these assumptions into attention, let us analyze evolutionary interrelations among different dialects of the genetic code from the viewpoint of the genetic octave Yin-Yang-algebra.

Only some triplets change their code meaning in the different dialects in comparison with the case of the vertebrate mitochondria code in the sense that they begin to encode other amino acids or stop-signal. What are those limitations which are utilized by the nature in its choice of such changeable (or evolutional) triplets? Has the matrix disposition of these variable triplets any relation to the YY-coordinates $x_0$, $x_1$, ..., $x_7$ and to their disposition in the genomatrix? Or the YY-coordinates have no relation to evolution of the genetic code and to systemic disposition of the variable triplets in the genomatrix $P^{(3)}$?

If such relation is discovered, it gives the additional evidence that the genetic octave Yin-Yang-algebra can be utilized as the adequate model of the genetic code or as the algebraic basis of the genetic code (the algebraic pre-code). It can be useful in tasks of sorting, putting in order and in deeper understanding of the genetic language. It can help to create new effective methods of information processing for many applied tasks as well. The appropriate algebraic model of the genetic code should give opportunities to deduce some evolutional peculiarities of the genetic code from such fundamental mathematical system.

The author's comparison analysis has discovered the expressed connection between the disposition of the variable triplets in the genomatrix $P^{(3)}$ and the disposition of the YY-coordinates $x_0$, $x_1$,.., $x_7$ together with their signs "+" and "-" in the matrix $YY_8$. The received results lead to a few phenomenological rules of evolution of the dialects of the genetic code on the basis of the genetic octave Yin-Yang-algebra. In other words the scheme, which is defined by this matrix algebra, holds true in the evolution of the genetic code in some significant aspects. These results give the additional evidence of appropriateness of such algebraic approach in bioinformatics.

Beginning with the level of the code correspondence between 64 triplets and 20 amino acids, some evolutional changes take place, which lead to many different dialects of the genetic code. Today the science knows 17 dialects of the genetic code, which differ by their numbers of degeneracy that is by the quantity of triplets encoding separate amino acids (Figure 4). Each amino acid is encoded in a concrete dialect by a certain quantity of triplets. This quantity of its triplets will be named "number of the degeneracy" of this amino acid in this dialect of the genetic code. For example, the amino acid Thr is encoded by 4 triplets in one genetic dialect; the number of degeneracy of this acid in this dialect is equal to 4. But this acid is encoded by 8 triplets in another dialect of the genetic code, where its number of degeneracy is equal to 8, etc. Structures of the set of such dialects reflect features of biological evolution on the very basic levels. It seems that the comparison analysis of these dialects can give important information about essence of biological organisms.

The modern science knows many dialects of the genetic code which exist in different kinds of organisms or of their subsystems (first of all, in mitochondria, which play a role of factories of energy in biological cells). For this article all initial data about the dialects of the genetic code were taken by the author from the NCBI's website http://www.ncbi.nlm.nih.gov/Taxonomy/Utils/wprintgc.cgi ). These dialects differ one from another through their specifics of the degeneracy (through concrete relations between 20 amino acids and 64 triplets). One can find from the data of the mentioned website, that 17 dialects are known only which differ one from another by the numbers of the degeneracy of the amino acids (see these 17 dialects in the table on Figure 4). A small quantity of the dialects from the website differ one from another by their start-codons only but not by the numbers of the degeneracy of the amino acids; we consider these dialects as the same dialect in our algebraic investigation. Let us analyze these 17 dialects of the genetic code from the viewpoint of the genetic octave Yin-Yang-algebra.

### 2.1. The numeric invariants of evolution and two branches of evolution in the set of dialects of the genetic code

The matrix form of presentation of members of the genetic octave Yin-Yang-algebra (Figure 2, on the right side) contains 32 components with the sign "+" and 32 components with the sign "-". The matrix disposition of the components with the sign "+"fits the disposition of the 32 black triplets. These black triplets encode 8 kinds of the high-degeneracy amino acids Ala, Arg, Gly, Leu, Pro, Ser, Thr, Val, each of which is encoded by 4 triplets or more in the vertebrate mitochondria genetic code, which is considered as the basic dialect [Petoukhov, 2001a, 2005a, arXiv:0803.3330]. Other 12 amino acids are encoded by the white triplets. These 12 acids Asn, Asp, Cys, Gln, Glu, His, Ile, Lys, Met, Phe, Trp, Tyr are the low-degeneracy ones because each of them is encoded by 3 triplets or less. So the set of 20 amino acids consists of the canonical sub-set of the 8 high-degeneracy amino acids and the canonical sub-set of the 12 low-degeneracy amino acids. In the case of the vertebrate mitochondria genetic code, the matrix disposition of these two sub-sets fits the matrix disposition of the YY-coordinates with the signs "+" and "-" correspondingly.

But do these two sub-sets, which fit the algebraic features of the matrix $YY_8$, play any role in many other dialects of the genetic code? The positive answer on this question is revealed, which leads to interesting phenomenological rules. In other words, the scheme, which is defined by this matrix algebra, is held true for the evolution of the genetic code in some significant aspects. The table on Figure 4 demonstrates 17 dialects of the genetic code with their numbers of degeneracy. Numbers of degeneracy, which are observed in the dialects, are equal to numbers from 1 to 8. For example, the first dialect of the genetic code in the table on Figure 4 possesses 12 amino acids, which number of degeneracy is equal to 2 (Asn, Asp, Cys, Gln, Glu, His, Ile, Lys, Met, Phe, Trp, Tyr); 6 amino acids, which number of degeneracy is equal to 4 (Ala, Arg, Gly, Pro, Thr,Val), and 2 amino acids, which number of degeneracy is equal to 6 (Leu, Ser). At first it seems, that the tabular distribution of numbers of degeneracy in a set of the 17 dialects is chaotic on the whole (Figure 4). But such impression disappears if one divides a series of all numbers of the degeneracy into two regular categories: the category of numbers from 1 to 3, which fits the low-degeneracy amino acids, and the category of numbers from 4 to 8, which fits to the high-degeneracy amino acids.

| The dialects of the genetic code | Distribution of numbers of degeneracy from 1 to 8 among 20 AA | | | | | | | | ΣAA with ND from **<u>1 to 3</u>** | ΣAA with ND from **<u>4 to 8</u>** |
|---|---|---|---|---|---|---|---|---|---|---|
| | 1 | 2 | 3 | 4 | 5 | 6 | 7 | 8 | | |
| 1) The Vertebrate Mitochondrial Code | | 12 | | 6 | | 2 | | | **12** | **8** |
| 2) The Standard Code | 2 | 9 | 1 | 5 | | 3 | | | **12** | **8** |
| 3) The Mold, Protozoan, and Coelenterate Mitochondrial Code and the Mycoplasma /Spiroplasma Code | 1 | 10 | 1 | 5 | | 3 | | | **12** | **8** |
| 4) The Invertebrate Mitochondrial Code | | 12 | | 6 | | 1 | | 1 | **12** | **8** |
| 5) The Echinoderm and Flatworm Mitochondrial Code | 2 | 8 | 2 | 6 | | 1 | | 1 | **12** | **8** |
| 6) The Euplotid Nuclear Code | 2 | 8 | 2 | 5 | | 3 | | | **12** | **8** |
| 7) The Bacterial and Plant Plastid Code | 2 | 9 | 1 | 5 | | 3 | | | **12** | **8** |
| 8) The Ascidian Mitochondrial Code | | 12 | | 5 | | 3 | | | **12** | **8** |
| 9) The Alternative Flatworm Mitochondrial Code | 2 | 7 | 3 | 6 | | 1 | | 1 | **12** | **8** |
| 10) Blepharisma Nuclear Code | 2 | 8 | 2 | 5 | | 3 | | | **12** | **8** |
| 11) Chlorophycean Mitochondrial Code | 2 | 9 | 1 | 5 | | 2 | 1 | | **12** | **8** |
| 12) Trematode Mitochondrial Code | 1 | 10 | 1 | 6 | | 1 | | 1 | **12** | **8** |
| 13) Scenedesmus obliq. mitochond. Code | 2 | 9 | 1 | 5 | 1 | 1 | 1 | | **12** | **8** |
| 14) Thraustochytrium Mitochondr. Code | 2 | 9 | 1 | 5 | 1 | 2 | | | **12** | **8** |
| 15) The Alternative Yeast Nuclear Code | 2 | 9 | 1 | 5 | 1 | 1 | 1 | | **12** | **8** |
| 16) The Yeast Mitochondrial Code | | 13 | | 5 | | 1 | | 1 | **13** | **7** |
| 17) The Ciliate, Dasycladacean and Hexamita Nuclear Code | 2 | 8 | 1 | 6 | | 3 | | | **11** | **9** |

Fig.4. The 17 dialects of the genetic code and distributions of their numbers of degeneracy (ND) among 20 amino acids (AA). The two right columns show quantities of the low-degenerate and high-degenerate acids (ΣAA). Bold frames mark two categories of numbers of the degeneracy: from 1 to 3 and from 4 to 8. Initial data were taken from the NCBI's website http://www.ncbi.nlm.nih.gov/Taxonomy/Utils/wprintgc.cgi.

From the two right columns of the table one can see the essential fact that the quantity of the low-degeneracy acids is equal to 12 and that the quantity of the high degeneracy acids is equal to 8 for different dialects. This fact, which is connected formally with the octave Ying-Yang algebra, can be formulated as the following rule, which has small exceptions only [Petoukhov, 2001a, 2005a].

**<u>The phenomenological rule № 1 connected with the octave Yin-Yang-algebra</u>**. In the set of the dialects of the genetic code, the set of 20 amino acids contains the two oppositional sub-sets: one sub-set consists of the 12 low-degeneracy acids (with the numbers of the degeneracy from 1 to 3) and the another sub-set consists of the 8 high-degeneracy acids (with the numbers of the degeneracy from 4 to 8). The sub-set of the 8 high-degeneracy acids together with their triplets corresponds to the YY-coordinates with the sign "+". The sub-set of the 12 low-degeneracy acids together with their triplets corresponds to the YY-coordinates with the sign "-".

As the author can judge, this rule about the canonical ratio 12:8 for two categories of amino acids is held true in nature without any exceptions for dialects of the genetic code of autotrophic organisms. (This type of organisms play the main role in biogeochemical cycles). But this rule has small exceptions in two cases of heterotrophic organisms in a form of minimal numeric shifting from the regular ratio "12 and 8" to the nearest integers ratios: The "Yeast Mitochondrial Code" possesses the ratio "13:7" for these two categories of amino acids, and the "Ciliate, Dasycladacean and Hexamita Nuclear Code" possesses the ratio "11 and 9" (see dialects 16 and 17 on Figure 4). These non-standard ratios encircle the canonical ratio "12:8" from the contrary sides of numeric axis. These non-standard ratios demonstrate additionally the main role of the canonical ratio 12:8 as that centre, around which minimal numeric fluctuations exist.

The data about evolution of the genetic code demonstrate the existence of the following rule about canonical sub-sets of the low-degeneracy and high-degeneracy amino acids also.

**<u>The phenomenological rule № 2 connected with the octave Yin-Yang-algebra.</u>** If a triplet encodes different amino acids in different dialects of the genetic codes, then these amino acids belong to the same canonical subset of amino acids of the low-degeneracy or high-degeneracy category. In other words, it is practically forbidden for those triplets, which encode amino acids of one canonical subset of degeneracy, to pass during biological evolution into the group of triplets, which encode amino acids of another canonical subset.

A single exception from this rule is the triplet UAG, which can encode amino acids Leu or Gln from different canonical subsets. The rule № 2 says nothing about stop-codons, and so it does not consider those evolutional cases, when triplets, which encode stop-codons (or amino acids) in one genetic code, begin to encode amino acids (or stop-codons respectively) in another code.

Described phenomenological rules testify that two independent branches of evolution of genetic code at billions biological objects exist: one branch exists for the canonical set of the high-degeneracy amino acids, and another branch exists for the canonical set of the low-degeneracy amino acids. These two independent branches have the relation to the two types of YY-coordinates with signs "+" and "-" from the proposed algebraic model $YY_8$ of the genetic code.

### 2.2 The conformity among YY-coordinates, triplets and amino acids for different dialects of the genetic code

As we mentioned above, only some triplets change their code meaning in the different dialects in comparison with the case of the vertebrate mitochondria code. What are those formal attributes which are utilized by the nature in its choice of these evolutional changeable triplets from the set of 64 triplets? How these triplets and their appropriate amino acids are disposed in the genomatrix $P^{(3)}$ (Figure 1) ? Has the matrix disposition of these variable triplets any relation to the YY-coordinates $x_0$, $x_1$, ..., $x_7$ and to their disposition in the genomatrix? Can these variable triplets be agreed naturally with the groups of the male and female YY-coordinates and triplets? Or the YY-coordinates have no relation to evolution of the genetic code and to systemic disposition of the variable triplets in the genomatrix $P^{(3)}$? This section continues the comparison analysis to answer such questions.

The table on Figure 5 gives data for such analysis. The vertebrate mitochondria genetic code (the code № 1) is utilized as the standard of comparison of code meanings of triplets in different dialects. The second tabular column shows those changeable triplets, which possess another code meaning (relative to their meaning in the dialect № 1) in the dialect which is named in the first column. A name of encoded amino acid or stop-codon (Stop) is given near each triplet in the second column in connection with the appropriate dialect named in the first column. Brackets in the second column contain that amino acid or stop-codon which is encoded by this triplet in the dialect № 1. Each row of the second column is finished by the YY-coordinate, which is disposed together with this triplet in the same cell of the matrix on Figure 1. At last, the third column demonstrates data about start-codons, which define the beginning of protein synthesis in the considered dialect. An appropriate YY-coordinate is shown for each start-codon as well.

| **Dialects of the genetic code** | **Changeable triplets** | **Start-codons** |
|---|---|---|
| 1) The Vertebrate Mitochondrial Code | | AUU, $-x_6$<br>AUC, $-x_6$<br>AUA, $-x_7$<br>AUG, $-x_7$<br>GUG, $x_3$ |
| 2) The Standart Code | UGA, Stop (Trp), $-x_5$<br>AGG, Arg (Stop), $-x_5$<br>AGA, Arg (Stop), $-x_5$<br>AUA, Ile (Met), $-x_7$ | UUG, $-x_7$<br>CUG, $x_3$<br>AUG, $-x_7$ |
| 3) The Mold, Protozoan, and Coelenterate Mitochondrial Code and the Mycoplasma/Spiroplasma Code | AGG, Arg (Stop), $-x_5$<br>AGA, Arg (Stop), $-x_5$<br>AUA, Ile (Met), $-x_7$ | UUG, $-x_7$<br>UUA, $-x_7$<br>CUG, $x_3$<br>AUC, $-x_6$<br>AUU, $-x_6$<br>AUG, $-x_7$<br>AUA, $-x_7$<br>GUG, $x_3$ |
| 4) The Invertebrate Mitochondrial Code | AGG, Ser (Stop), $-x_5$<br>AGA, Ser (Stop), $-x_5$ | UUG, $-x_7$<br>AUU, $-x_6$<br>AUC, $-x_6$<br>AUA, $-x_7$<br>AUG, $-x_7$<br>GUG, $x_3$ |
| 5) The Echinoderm and Flatworm Mitochondrial Code | AGG, Ser (Stop), $-x_5$<br>AGA, Ser (Stop), $-x_5$<br>AUA, Ile (Met), $-x_7$<br>AAA, Asn (Lys), $-x_7$ | AUG, $-x_7$<br>GUG, $x_3$ |
| 6) The Euplotid Nuclear Code | UGA, Cys (Trp), $-x_5$<br>AGG, Arg (Stop), $-x_5$<br>AGA, Arg (Stop), $-x_5$<br>AUA, Ile (Met), $-x_7$ | AUG, $-x_7$ |
| 7) The Bacterial and Plant Plastid Code | UGA, Stop (Trp), $-x_5$<br>AGG, Arg (Stop), $-x_5$<br>AGA, Arg (Stop), $-x_5$<br>AUA, Ile (Met), $-x_7$ | UUG, $-x_7$<br>CUG, $x_3$<br>AUC, $-x_6$<br>AUU, $-x_6$<br>AUA, $-x_7$<br>AUG, $-x_7$ |
| 8) The Ascidian Mitochondrial Code | AGG, Gly (Stop), $-x_5$<br>AGA, Gly (Stop), $-x_5$ | UUG, $-x_7$<br>AUA, $-x_7$<br>AUG, $-x_7$<br>GUG, $x_3$ |
| 9) The Alternative Flatworm Mitochondrial Code | UAA, Tyr (Stop), $-x_7$<br>AGG, Ser (Stop), $-x_5$<br>AGA, Ser (Stop), $-x_5$<br>AUA, Ile (Met), $-x_7$<br>AAA, Asn (Lys), $-x_7$ | AUG, $-x_7$ |
| 10) Blepharisma Nuclear Code | UGA, Stop (Trp), $-x_5$<br>UAG, Gln (Stop), $-x_7$<br>AGG, Arg (Stop), $-x_5$<br>AGA, Arg (Stop), $-x_5$<br>AUA, Ile (Met), $-x_7$ | AUG, $-x_7$ |
| 11) Chlorophycean Mitochondrial Code | UGA, Stop (Trp), $-x_5$<br>UAG, Leu (Stop), $-x_7$<br>AGG, Arg (Stop), $-x_5$<br>AGA, Arg (Stop), $-x_5$<br>AUA, Ile (Met), $-x_7$ | AUG, $-x_7$ |
| 12) Trematode Mitochondrial Code | AGG, Ser (Stop), $-x_5$<br>AGA, Ser (Stop), $-x_5$<br>AAA, Asn (Lys), $-x_7$ | AUG, $-x_7$<br>GUG, $x_3$ |
| 13) Scenedesmus obliquus Mitochondrial Code | UGA, Stop (Trp), $-x_5$<br>UAG, Leu (Stop), $-x_7$<br>UCA, Stop (Ser), $x_5$<br>AGG, Arg (Stop), $-x_5$<br>AGA, Arg (Stop), $-x_5$<br>AUA, Ile (Met), $-x_7$ | AUG, $-x_7$ |
| 14) Thraustochytrium Mitochondrial Code | UGA, Stop (Trp), $-x_5$<br>UUA, Stop (Leu), $-x_7$<br>AGG, Arg (Stop), $-x_5$<br>AGA, Arg (Stop), $-x_5$<br>AUA, Ile (Met), $-x_7$ | AUU, $-x_6$<br>AUG, $-x_7$<br>GUG, $x_3$ |
| 15) The Alternative Yeast Nuclear Code | UGA, Stop (Trp), $-x_5$<br>AGG, Arg (Stop), $-x_5$<br>AGA, Arg (Stop), $-x_5$<br>AUA, Ile (Met), $-x_7$<br>CUG, Ser (Leu), $x_3$ | CUG, $x_3$<br>AUG, $-x_7$ |
| 16) The Yeast Mitochondrial Code | AGG, Arg (Stop), $-x_5$<br>AGA, Arg (Stop), $-x_5$<br>CUG, Thr (Leu), $x_3$<br>CUU, Thr (Leu), $x_2$<br>CUA, Thr (Leu), $x_3$<br>CUC, Thr (Leu), $x_2$ | AUA, $-x_7$<br>AUG, $-x_7$ |
| | | |
| 17) The Ciliate, Dasycladacean and Hexamita Nuclear Code | UGA, Stop (Trp), **-$x_5$**<br>UAG, Gln (Stop), **-$x_7$**<br>UAA, Gln (Stop), **-$x_7$**<br>AGG, Arg (Stop), **-$x_5$**<br>AGA, Arg (Stop), **-$x_5$**<br>AUA, Ile (Met), **-$x_7$** | AUG, **-$x_7$** |

Figure 5. The table about changeable triplets and start-codons in the dialects of the genetic code. Initial data are taken from http://www.ncbi.nlm.nih.gov/Taxonomy/Utils/wprintgc.cgi.

**About triplets which change their code meaning**. Let us analyze the data from the second column of the table on Figure 5. This column shows 14 kinds of the changeable triplets which possess different code meanings in different dialects: AAA, AGA, AGG, AUA, CUA, CUC, CUG, CUG, CUU, UAA, UAG, UCA, UGA, UUA. Some of these triplets have several meanings. For example the triplet AGA encodes the stop-signal in the dialect № 1, the amino acid Arg in the dialect № 4; the amino acid Gly in the dialect № 8. Or the triplet UAA encodes the stop-signal in the dialect № 1, the amino acid Tyr in the dialect № 9, the amino acid Gln in the dialect № 17.

All kinds of changeable triplets are met 69 times in the second column. But only two kinds of the male YY-coordinates "-$x_5$" and "-$x_7$" with the sign "-" correspond to these triplets in all dialects practically. Specifically the male coordinate "-$x_5$" is met 41 times (it is 59,4% of all cases), and the male coordinate "-$x_7$"is met 22 times (it is 31,9% of all cases). It composes in sum more than 90% of all cases. The male coordinate "+$x_5$" is met 1 time in the dialect № 13 but with the sign "+". One can name the male YY-coordinates "-$x_5$", "-$x_7$" and "+$x_5$" as canonical Yin-Yang-coordinates for the changeable triplets (Figure 1). The described statistics permits to formulate the following rule.

**The phenomenological rule № 3 connected with the octave Yin-Yang-algebra**. Those triplets possess different code meanings in the different dialects of the genetic code, which correspond to the canonical male coordinates "-$x_5$", "-$x_7$" and "+$x_5$" of the matrix $YY_8$.

This rule is held true precisely for all the dialects besides the case of yeast with their two dialects: the dialect № 15, where the non-canonical male coordinate "+$x_3$" appears (for the triplet CUG), and the dialect № 16, which has the following unique feature. In this dialect № 16 the four triplets CUA, CUG, CUC, CUU, which are begun with the same pair of the letters (CU), change their code meanings by the identical way: all of them encode Thr instead of Leu (it is unusual case because, if any other four triplets are begun with the equal pair of any letters, they do not change jointly their code meanings in other dialects). These four triplets correspond to the non-canonical YY-coordinates "+$x_2$" and "+$x_3$".

Yeast is unicellular mushrooms, chemoorganoheterotrophs, which are possible to vegetative cloning (asexual reproduction). Probably, the genetic-code deviation of the yeast from the rule № 3 is connected with their asexual reproduction and heterotrophy. (We noted in the section 2.1 already, that the dialects of the genetic code of the heterotrophs, which feed on ready living substance, can have some deviations from the canonical forms of the dialects of autotrophic organisms). The additional evidence of molecular-genetic singularity of yeast is the fact that the histone H1 is not discovered in their genetic system at all (http://drosophila.narod.ru/Review/histone.html).

**The connection between evolution of the genetic code and the anisotropy of the $YY_8$-space**. The previous author's article [Petoukhov, arXiv:0803.3330, version 2, sections 5-7] has described the anisotropy of the coordinate space of the $YY_8$-numbers (the $YY_8$-space). The 8-dimensional $YY_8$-numbers $YY_8 = x_0*\mathbf{f_0}+x_1*\mathbf{m_1}+x_2*\mathbf{f_2}+x_3*\mathbf{m_3}+x_4*\mathbf{f_4}+x_5*\mathbf{m_5}+x_6*\mathbf{f_6}+x_7*\mathbf{m_7}$ have been interpreted as the double genetic quaternion. If all female coordinates $x_0$, $x_2$, $x_4$, $x_6$ are equal to 0, we have the male variant of $YY_8$:

$$(YY_8)_{MALE} = x_1*\mathbf{m_1} + x_3*\mathbf{m_3} + x_5*\mathbf{m_5} + x_7*\mathbf{m_7} \quad (2)$$

The table of multiplication of the basic elements $\mathbf{m_1}$, $\mathbf{m_3}$, $\mathbf{m_5}$, $\mathbf{m_7}$ of $(YY_8)_{MALE}$ coincides with the multiplication table of genetic quaternions $g = y_0*\mathbf{1} + y_1*\mathbf{i_1} + y_2*\mathbf{i_2} + y_3*\mathbf{i_3}$ (see [Petoukhov, arXiv:0803.3330, version 2, expression 9 and Figure 14]). By analogy with Hamilton's quaternion, the first item $y_0*\mathbf{1}$ of genetic quaternions can be named their scalar part and the sum of other three items can be named the vector part of genetic quaternions.

In accordance with Figure 15 in [Petoukhov, arXiv:0803.3330, version 2] the number (like other genetic quaternions) possesses the norm

$$|(YY_8)_{MALE}|^2 = x_1^2 + x_3^2 - x_5^2 - x_7^2 \quad (3)$$

The signature (+,+,-,-) of the norm (3) of genetic quaternions differs from the signature (+,+,+,+) of the norm of quaternions by Hamilton. This difference is very significant because it defines the following fundamental circumstance. The vector part $x_3*\mathbf{m_3} + x_5*\mathbf{m_5} + x_7*\mathbf{m_7}$ of genetic quaternions corresponds to the case of anisotropic space in contrast to quaternions by Hamilton, the vector part of which corresponds to the case of the isotropic space [Petoukhov, arXiv:0803.3330, version 2, sections 5-7]. This difference in the signatures of the norms is connected in the expression (3) with the YY-coordinates $x_5$ and $x_7$, which can be named "anisotropic coordinates" in this reason. But these coordinates $x_5$ and $x_7$ are those, which correspond to the changeable triplets of the genetic code in accordance with the rule № 3. It is very interesting fact that all evolution of code meanings of genetic triplets occurs in connection with these anisotropic coordinates of the model space practically. Consequently the close connection between evolution of the genetic code and the anisotropy of this $YY_8$-space exists. In this reason, one can formulate the following rule № 4, which is a continuation of the rule № 3.

**The phenomenological rule № 4, which is connected with the genetic octave Yin-Yang-numbers and with the anisotropy of $YY_8$-space**. In evolution of dialects of the genetic code, all changeable triplets correspond to the anisotropic male coordinates of genetic $YY_8$-numbers.

Similarly to the rule № 3, the rule № 4 has one exception: the case of yeast, which is characterized by asexual reproduction and heterotrophy and which changes the code meanings of the coordinates $x_2$ and $x_3$ additionally. It is obvious that the following **prediction** can be made. If new dialects of the genetic code will be discovered in the future for organisms with bisexual reproduction, changeable triplets there will correspond to the anisotropic male coordinates of genetic $YY_8$-numbers as well.

One can make one more remark about the male coordinates "$-x_5$", "$-x_7$", which are connected with more than 90% of all changeable triplets, as it was mentioned above. All triplets, which correspond to these coordinates, change their code meanings besides the four invariable triplets: UGG with the coordinate "$-x_5$", and AAG, AUG, UUG with the coordinate "$-x_7$". Perhaps new dialects of the genetic code will be discovered in the future, where these triplets change their code meanings as well.

**The phenomenologic rule № 5 connected with the genetic octave Yin-Yang-numbers.** All those 16 triplets, which correspond to the YY-coordinate $x_0$ and $x_1$ of the scalar part of genetic $YY_8$-numbers, never change their code meanings (including the case of yeast).

Really, one can see that these coordinates $x_0$ and $x_1$ of the scalar part of $YY_8$-numbers are absent in the table on Figure 5 together with their 16 triplets CCC, CCA, CCU, CCG, CGC, CGA, CGU, CGG, GCC, GCA, GCU, GCG, GGC, GGA, GGU, GGG. So, the coordinates of the scalar part of the genetic $YY_8$-numbers define the absolute invariable part of the set of genetic triplets.

**About stop-codons**. Encoding of stop-signals of protein synthesis turns on a special interest. Stop-signals are encoded by different triplets (stop-codons) in different dialects of the genetic code. The 7 kinds of triplets play the role of stop-codons in these dialects. Three of them (UUU, UAG, UUA) fit the YY-coordinate "-$x_7$". Other three triplets (AGA, AGG, UGA) fit the coordinate "-$x_5$". The seventh triplet (UCA) fits the coordinate "+$x_5$". All these coordinates are the anisotropic male YY-coordinates. Consequently the function of stop-codons is close connected with the anisotropy of $YY_8$-space. The results of the investigation of stop-codons in the dialects from the viewpoint of $YY_8$-algebra permit to formulate the following rule.

**The phenomenological rule № 6 connected with the octave Yin-Yang-algebra and with the anisotropy of the $YY_8$-space**.Those triplets serve as stop-codons in the dialects of the genetic code, which correspond to the anisotropic male YY-coordinates "-$x_5$",."-$x_7$" and "+$x_5$".

This rule is held true for all 17 dialects without exceptions. It pays attention to the fact that the function of stop-codons is the "male function" always from the viewpoint of $YY_8$-algebra because stop-codons are connected with the male coordinates. A few triplets exist (for example UUA and UGG), which correspond to the same coordinates «-$x_5$», «-$x_7$» и «+$x_5$» but which are not stop-codons in known dialects of the genetic code. Whether such dialect of the genetic code will discover in the future, where these triplets play the role of stop-codons? The time will show.

**About start-codons**. Till now we did not analyze start-codons (function of start-codons is the additional function of some triplets which they execute besides their basic function of coding of amino acids). The third column of the table on Figure 5 demonstrates those start-codons of the 17 dialects of the genetic code, which are presented in basic sets of code meanings of 64 triplets of the considered 17 dialects on the website http://www.ncbi.nlm.nih.gov/Taxonomy/Utils/wprintgc.cgi. Eight triplets play the role of start-codons in these 17 cases. The four of them (AUA, AUG, UUA, UUG) correspond to the YY-coordinate "-$x_7$". The two triplets (AUC, AUU) correspond to the coordinate "-$x_6$". Other two triplets (CUG, GUG) correspond to the coordinate "+$x_3$". The set of start-codons of the dialect № 1 corresponds to all these coordinates "-$x_7$", "-$x_6$" and "+$x_3$". These data permit to formulate the additional rule about start-codons.

**The phenomenological rule № 7 connected with the octave Yin-Yang-algebra**. All start-codons in the dialects of the genetic code correspond to YY-coordinates "-$x_7$", "-$x_6$" and "+$x_3$".

This rule is held true for all 17 dialects of the genetic code without exceptions. One can add that the start-codon AUG, which corresponds to the YY-coordinate "-$x_7$", is included in all the 17 dialects. All start-codon, which are presented in the table on Figure 5, have the letter U on their second position that reminds about the U-algorithm of connection between genomatrices and Hadamard matrices [Petoukhov, arXiv:0802.3366; arXiv:0803.0888; arXiv:0803.3330].

## 3 The molecular-sexual approach in molecular genetics

Let us name each low-degeneracy amino acid, which is encoded by one of the female YY-coordinates $x_0$, $x_2$, $x_4$, $x_6$ only (Figure 1), as the female amino acid conditionally. Such female amino acids are Asn, Asp, Cys, His, Ile, Phe, Tyr (we mark the female acids by pink color by analogy with the female YY-coordinates). Each low-degeneracy amino acid, which is encoded by one of the male YY-coordinates $x_1$, $x_3$, $x_5$, $x_7$ only, is named as the male amino acid correspondingly. Such male amino acids are Gln, Glu, Lys, Met, Trp (we mark the male acids by blue color).

The case of the high-degeneracy amino acids is more complex because such acids correspond to male and female YY-coordinates simultaneously (Figure 1). For example the acid Arg corresponds to $x_0$ and $x_1$. In this reason we name each of the high-degeneracy amino acids as the androgynous acid conditionally by analogy with androgynous persons which possess male and female attributes simultaneously. The pure androgynous acids, each of which corresponds to the male and female YY-coordinates in the equal degree, are Ala, Arg, Gly, Pro, Thr, Val (we mark androgynous acids by green color). The amino acid Ser is disposed in 6 cells in the octave genomatrix (Figure 1), which correspond to the unequal quantities of the female and male YY-coordinates: the quantity of the female coordinates is equal to 4 and the quantity of the male coordinates is equal to 2. In this reason Ser is named the “androgynous acid of the female type”. The amino acid Leu possess the symmetric-oppositional character relative to Ser because Leu is disposed in 6 matrix cells also but these cells correspond to 4 male coordinates and to 2 female coordinates (Figure 1). In this reason Leu is named the “androgynous acid of the male type”. Leu and Ser form the sub-set of the quasi-androgynous acids.

All these names are introduced on the basis of the vertebrate mitochondria code (the code № 1 on Figures 4 and 5). But what we have for the other 16 dialects of the genetic code, where some amino acids receive new correspondence to triplets and YY-coordinates? What such changes mean from the viewpoint of the notions of the genetic $YY_8$-algebra?

In accordance with the rule № 3, some male triplets of the dialect № 1 change their code meanings in the course of evolution of the genetic code (the case of yeast is the exceptional one). These male triplets encode the male and androgynous acids in the dialect № 1. The additional rule is that, if such changeable male triplet encodes another amino acid in another dialect of the genetic code, this new amino acid is one of the female acids necessarily (the case of yeast is the exceptional one). In other words, the expansion of the female acids (Asn, Cys, Ile, Tyr) into the male columns of the genomatrix $P^{(3)}$ (Figure 1) take place in the course of evolution. But the male amino acids never come to the female columns. In the result the genomatrix $P^{(3)}$, which possesses the equal qualities of the male and female amino acids in the dialect № 1, becomes the more female matrix in other dialects due to prevalence of the female amino acids in the matrix cells there.

One can add that the androgynous acids Arg, Gly, Leu force out the male stop-codons in some dialects and take their places in the male columns of the genomatrix $P^{(3)}$. Figuratively speaking, the female beginning forces out the male beginning in the set of amino acids. On the other hand, the male triplets increase their positions in the set of start-codons and stop-codons to guide punctuations of protein synthesis.

They encode not only all stop-codons in all dialects, but the set of start-codons becomes the more male set in the course of evolution: the single female coordinate “$-x_6$”, which exists in the dialect № 1, is eliminated in the most dialects. Really the dialects №№ 2, 5, 6, 8-13, 15-17 have not start-codons with the female YY-coordinates (Figure 5). On the whole the evolution of the genetic code is the struggle between the male and female beginnings on the molecular-genetic level from the viewpoint of this algebraic model. It reminds of the struggle between matriarchy and patriarchy in the history of human civilization. It reminds of many other famous confrontations between the male and female beginnings as well. The creator of analytic psychology C.Yung subdivides the soul into the male and female beginnings, ratios of which can be changed in different periods of human life. The described data permits to formulate the following rule of the struggle between male and female beginnings in evolution of the genetic code from the viewpoint of the genetic $YY_8$-algebra.

**The phenomenological rule № 8 connected with the octave Yin-Yang-algebra**. In evolution of dialects of the genetic code, an increase of the set of triplets, which encode the female and androgynous amino acids, exists concerning the analogical set in vertebrate mitochondria genetic code. The set of triplets, which encode the start-codons, become the more male set in this process.

The revealing such structural division of the set of 20 amino acids into the sub-sets of the male, female and androgynous amino acids can be useful for modeling many astonishing phenomena in molecular genetics. The speech is about the phenomena of mutual disclosure and of mutual attraction between two molecular one-specific partners, which lead to formation of new molecular pairs; they take place in a medium of huge number of other molecules (molecular bouillon). Let us consider the following example.

### 3.1 The example of the pairs of histones

It is known that nucleosome histones are important protein components of chromosomes. We utilize the well-known data about histones from the NCBI's websitehttp://www.ncbi.nlm.nih.gov/books/bv.fcgi?rid=mboc4.figgrp.632 and from the website of the Institute of the molecular biology of the Russian Academy of Sciences http://articles.excelion.ru/science/biology/22569393.html (the article by V.L.Karpov).

Filaments of DNA in eukaryote cells are coiled around nucleosomes, each of which is a shank consisted of the histones of the four types: H2A, H2B, H3 and H4. This set of four types is divided by the nature into the pairs of one-specific histones. The histones H2A and H2B possess the important possibility to create the pair just one with another on the basis of their mutual revealing and mutual "attraction" in molecular bouillon (by analogy with a male and a female of one species among macroscopic biological organisms). Another pair consists of the histones H3 and H4, which possess the similar possibility to create the pairs just one with another on the analogical basis of their mutual revealing and mutual "attraction" in molecular bouillon.

A single nucleosome contains the ensemble of eight histones, where two histones of each of the four types H2A, H2B, H3 and H4 are included. The DNA molecule is reeled up on this octamer shank in the form of the left spiral. The structure of nucleosome plays the main role in packing of DNA on all levels. Each nucleosome is formed in accordance with the principle of the multi-level recognition defined by the structures of the histones. Each histone molecule contains a central structured 3-spiral domain and non-structured N- and C-"tails". The one-specific histones identify one another and create their pairs. All creation of the octamer shank is based on the consecutive creation of pairs of the two one-specific molecular objects (Figure 6).

On the first step spiral domains cooperate among themselves. In the result, pairs (dimers) arise: one pair H3-H4 and two pairs H2A-H2B. On the second step, two first dimers form the pair association of the following level of complexity: the tetramer with two pairs H3-H4. On the third step this tetramer forms a pair association of the higher level with two pairs H2A-H2B. In the result, the octamer of the histones arises. All these searches and copulations one-specific histones into pairs, and then into new pairs from pairs occur in a molecular bouillon with a huge bedlam of biomolecules of other kinds and their splinters. It occurs despite of effects of electric shielding and other noise circumstances there.

These phenomena of micro world of molecules should be subordinated to principles of quantum mechanics. But their conclusion from these known principles is an excessive problem for the modern science. These paired associations of the histones in molecular genetics carry the art name "hand shakes of molecules" traditionally (Figure 6). But from the viewpoint of the our bisex theory, which is based on the genetic Yin-Yang algebras and which says about "sexual" interactions of genetic molecules, one can propose another art name for such pair search and association: "a family", "a love-crossing " or "a love-copulation "

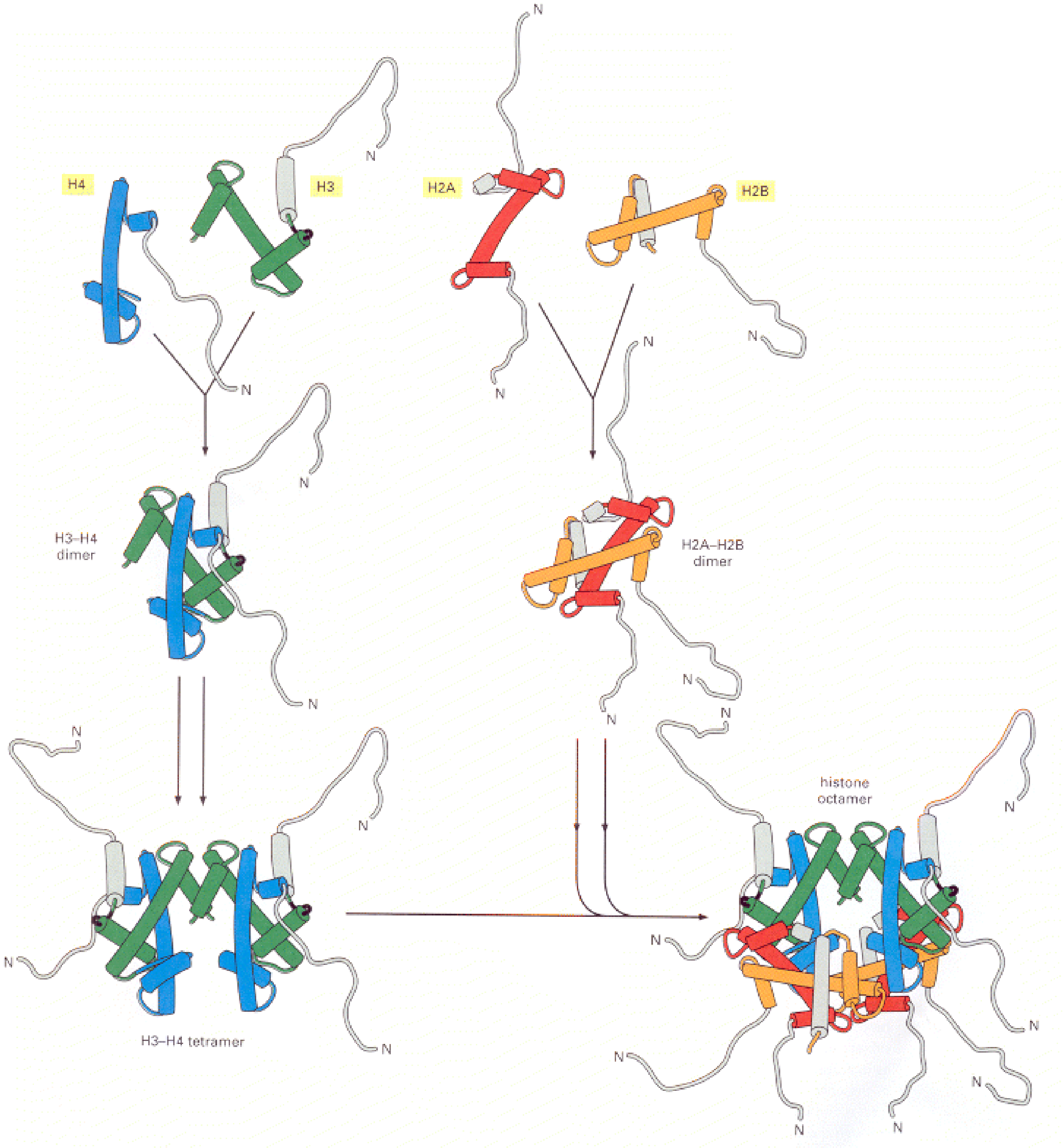

Figure 6. The multistage association of pairs of one-specific histones H3-H4 and H2A-H2B into the dimers, tetramers and octamers (this figure is taken from http://www.ncbi.nlm.nih.gov/books/bv.fcgi?rid=mboc4.figgrp.636)

### 3.2 Whether an unknown quantum mechanical factor of a “sexual attraction” among genetic molecules exists?

Such molecular-genetic facts give the basis to suspect, that phenomena of love or love search of the sexual partner, which exist at a level of animal organisms, have arisen not on an empty place. But they are the continuation of quantum mechanical phenomena of search of the one-specific pair partner, which exist already at the level of genetic molecules, at least. The interrelations among genetic molecules in their search of each other can be interpreted as sexual relations to some extent.

Plato had formulated the famous statement about a congenital aspiration of each person to look for the second half. From the viewpoint of the our bisex conception, which is based on the genetic Yin-Yang algebra, the Plato's statement can be transferred into the world of those congenital properties of genetic molecules which are reflected in their search of their second halves.

In our opinion, taking into account the described facts, one can put forward the working hypothesis about existence of "a sexual intermolecular attraction" (or a "bisex attraction") between genetic one-specific elements as new biophysical factor of a quantum mechanical sense. This new hypothetical factor or principle is entered, first of all, for an explanation of molecular-genetic phenomena of search of the one-specific pair partner by multi-atomic biomolecules to create a specific pair in complex conditions of multi-component bullion. The genetic Yin-Yang-algebra can be useful to model and investigate such factor. This factor can have a force character and/or information character. It does not reject existence of other known factors (for example, interactions of electric charges and so forth), but it is additional to them. Of course, it would be wrong to extend an action of this factor of "a sexual intermolecular attraction", which is proposed in connection with phenomena of assembly of pairs of one-specific multi-atomic molecular elements (multi-atomic quantum mechanical “modules”), into the field of all aspects of molecular-genetic organization.

Another example of possible display of the hypothetical factor of "a sexual intermolecular attraction" in genetic systems gives the phenomenon which was discovered by Mirzabekov at studying a transfer RNA [Mirzabekov, 1997]. The speech is that halves and quarters of these molecules can find each other in molecular bouillon and gather in one molecule, which possesses a typical function of a transfer RNA.

### 3.3 The analysis of the insulin structure as the example of the bisex analysis of proteins

The discovered connection between the genetic code and the Yin-Yang-algebra gives the opportunity of classification of many molecular elements in accordance with their “sexual” (bisex) characteristics. The knowledge about such hidden structure of the set of genetic molecules can be useful for studing, an explanation and a prediction of features of interactions of molecular elements with different sexual characteristics.

For example, all set of proteins can be divided into sub-sets of male, female and androgynous proteins conditionally in accordance with their amino acids compositions. In the result new information arises about structural characteristics of proteins. Let us make the first attempt to analyze a protein from the viewpoint of the genetic Yin-Yang-algebra. For this first attempt we take the simplest protein – insulin, which contains 51 amino acids. The set of these acids are encoded by 51 triplets in the gene of insulin. Whether this genetic sequence of triplets has any regularity from the viewpoint of the genetic Yin-Yang algebra?

The insulin consists of two chains: the α-chain and the β-chain. The α-chain contains 21 amino acids and the β-chain contains 30 amino acids. These chains are encoded by the genetic sequences (Figure 7), which are taken from the text-book [Inge-Vechtomov, 1983, p. 321-323]. The intrinsic "sexual" structure of each sequence is demonstrated by appropriate colors for male (blue), female (pink) and androgynous (green) elements.

| | |
|---|---|
| α-chain | 1GGC(Gly, $x_0$)→2ATC(Ile,$-x_6$)→3GTT(Val,$x_2$)→4GAA(Glu,$-x_3$)→<br>5CAG(Gln,$-x_3$) →6TGT(Cys,$-x_4$)→7TGC(Cys,$-x_4$)→8ACT(Thr,$x_4$)→9TCT(Ser,$x_4$)→<br>10ATC(Ile,$-x_6$)→11TGC(Cys,$-x_4$)→12TCT(Ser,$x_4$)→13CTT(Leu,$x_2$)→<br>14TAC(Tyr,$-x_6$)→15CAG(Gln,$-x_3$)→16CTT(Leu,$x_2$)→17GAG(Glu,$-x_3$)→<br>18AAC(Asn,$-x_6$)→9TAC(Tyr,$-x_6$)→20TGT(Cys,$-x_4$)→21AAC(Asn,$-x_6$) |
| β-chain | 1TTC(Phe,$-x_6$)→2GTC(Val,$x_2$)→3AAT(Asn,$-x_6$)→4CAG(Gln,$-x_3$)→5CAC(His,$-x_2$)<br>→6CTT(Leu,$x_2$)→7TGT(Cys,$-x_4$)→8GGT(Gly,$x_0$)→9TCT(Ser,$x_4$)→10CAC(His,$-x_2$)<br>→11CTC(Leu,$x_2$)→12GTT(Val,$x_2$)→13GAA(Glu,$-x_3$)→14GCT(Ala,$x_0$)→<br>15TTG(Leu,$-x_7$)→16TAC(Tyr,$-x_6$)→17CTT(Leu,$x_2$)→18GTT(Val,$x_2$)→<br>19TGC(Cys,$-x_4$)→20GGT(Gly,$x_0$)→21GAA(Glu,$-x_3$)→22CGT(Arg,$x_0$)→<br>23GGT(Gly,$x_0$)→24TTC(Phe,$-x_6$)→25TTC(Phe,$-x_6$)→26TAC(Tyr,$-x_6$)→<br>27ACT(Thr,$x_4$) →28CCT(Pro,$x_0$)→29AAG(Lys,$-x_7$)→30ACT(Thr,$x_4$) |

Figure 7. The genetic sequences of triplets for the α-chain and the β-chain of insulin. The following data are shown for each triplet: its current number in the sequence; the encoded amino acid; its YY-coordinate from Figure 1. The male triplets and amino acids are marked by blue color; the female triplets and acids – by pink color; the androgynous triplets and acids – by green color (see the beginning of the section 3). The genetic letter T (thymine) is used instead of the letter U, but it does not matter.

The first results of this analysis are the following. The genetic sequences of the both chains contain the equal quantities of the female triplets (10 female triplets) and of the male triplets (4 male triplets). All other triplets are androgynous ones. From the described algebraic viewpoint, insulin and its gene are the female protein and the female gene because female acids and female triplets dominate there.

The α-chain has the following set of the female links: 2ATC(Ile,$-x_6$), 6TGT(Cys,$-x_4$), 7TGC(Cys,$-x_4$) , 10ATC(Ile,$-x_6$), 11TGC(Cys,$-x_4$), 14TAC(Tyr,$-x_6$), 18AAC(Asn,$-x_6$) , 19TAC(Tyr,$-x_6$), 20TGT(Cys,$-x_4$), 21AAC(Asn,$-x_6$). This set of the female links shows unexpectedly the regularity of double composition in its structure (the phenomenon of doubling).

The essence of this phenomenon is that each kind of female triplets and amino acids is met twice in the α-chain exactly. (Below we note that the analogical phenomenon of doubling takes place for the set of the male links also). Really, the female triplet ATC exists there in the links 2 and 10; the triplet TGT – in the links 6 and 20; the triplet TGC – in the links 7 and 11; the triplet TAC – in the links 14 and 19; the triplet AAC – in the links 18 and 21.

Let us make a small deviation into the field of linguistics. It is well-known in molecular genetics that "the more we understand laws of coding of genetic information, the more their similarity to principles of linguistics amazes" [Ratner, 2002, p.203].

Whether the described phenomenon of doubling in the case of the simplest protein (insulin) has a similarity with the structural principles of simplest words of human languages? Yes, it has. Really the simplest words of different human languages demonstrate the same phenomenon of doubling or of their construction on the basis of doubling letters. Such words are utilized be babies, when they start to speak; they are used intuitively by mothers in dialogue with babies; they are most digestible and exploitable at training speech in the case of deaf-and-dumb people; they are utilized by different nations for speech imitation of sounds of world around: “mama”, “papa”, “baba”, “wee-wee”, etc. (the Russian language, which is native for the author, has a lot of examples of such simplest words with doubling letters). In process of development of speech, this primitive principle of construction of the simplest words with doubling letters is overcome gradually. These data are the addition to the famous conception that linguistic languages have arisen not on an empty space but they are a continuation of the genetic language [Baily, 1982; Jacob, 1974, 1977; Makovskiy, 1992; Petoukhov, 2003, 2004; Jacobson, 1987, 1999; Yam, 1995; etc].

One small addition can be made else to this conception. The theory of artificial and computer languages demonstrates that there is no necessity at all to include in languages a division of the whole set of nouns (and some other language elements) into sub-sets of nouns of masculine gender, of feminine gender and of neuter gender. But the natural human languages possess such division of the set of all nouns into sub-sets of nouns of such three genders. If the human languages are continuation of the genetic language [Jacobson, 1987, 1999], then the genetic language should possess such division of the whole set of its elements into sub-sets of masculine, feminine and neuter genders. Our algebraic investigation of the genetic code confirms this conception by means of the discovery of the algebraic division of the sets of elements of the genetic languages into sub-sets of elements of masculine, feminine and neuter (androgynous) genders. In our opinion, these three genders of elements in genetic systems exist due to intrinsic features of the genetic Yin-Yang-algebra in connection with the fundamental task of noise immunity of genetic coding.

Let us return to insulin. The set of the female links of the α-chain contains 5 different triplets ATC, TGT, TGC, TAC, AAC, but 4 different amino acids exist there because the acid Cys exists in two pairs of links. Each of these pairs is encoded by its own triplet – TGT or TGC, but these triplets do not differ by their coding meanings because they encode the same amino acid Cys in all dialects of the genetic code. All this set of the female links corresponds to two YY-coordinates “-$x_6$” and “-$x_4$” only. The number of repetitions of each of these coordinates is even number: “-$x_6$” is repeated 6 times, and “-$x_4$” is repeated 4 times. One can remind that all YY-coordinates have the one-to-one relation with the letter composition of triplets [Petoukhov, arXiv:0803.3330, version 2, Figure 3]. In accordance with this connection, each abstract YY-coordinate (for example “-$x_6$”) presents an algorithmically defined number (“-ββγ” in this example), which is based on real parameters of the molecules of the genetic alphabet.

The male links of the α-chain are 4GAA(Glu,-$x_3$), 5CAG(Gln,-$x_3$), 15CAG(Gln,-$x_3$), 17GAG(Glu,-$x_3$). Each of their amino acids Glu and Gln is repeated twice again. All these links correspond to the same YY-coordinate “-$x_3$”. The triplet CAG is repeated twice. The triplets GAA and GAG do not differ in their coding meanings because they encode the same amino acid Gln in all dialects of the genetic code. So the phenomenon of doubling exists for the male links of the α-chain as well.

Each of the quasi-androgynous acids Ser and Leu exists in the α-chain twice also: 9TCT(Ser, $x_4$), 12TCT(Ser, $x_4$), 13CTT(Leu, $x_2$), 16CTT(Leu, $x_2$). These acids and their triplets TCT and CTT correspond to the female YY-coordinates “$x_2$” and “$x_4$”.

Now let us consider the β-chain of insulin (Figure 7) with the following female links: 1TTC(Phe,-$x_6$), 3AAT(Asn,-$x_6$), 5CAC(His,-$x_2$), 7TGT(Cys,-$x_4$), 10CAC(His,-$x_2$), 16TAC(Tyr,-$x_6$), 19TGC(Cys,-$x_4$), 24TTC(Phe,-$x_6$), 25TTC(Phe,-$x_6$), 26TAC(Tyr,-$x_6$). The phenomenon of doubling exists for the whole β-chain except of its first three links Phe-Val-Asn (which correspond to the tripeptide). Really the main part of the β-chain contains the following amino acids twice: His (encoded by the triplet CAC); Tyr (encoded by TAC); Phe (encoded by TTC); Cys (encoded by TGT and TGC, which do not differ in their code meanings in all dialects of the genetic code because they encode the same acid Cys always). The female links in the β-chain correspond to the YY-coordinates "-$x_6$", "-$x_4$" and "-$x_2$".

The set of the male links in the β-chain contains 4CAG(Gln, -$x_3$), 13GAA(Glu, -$x_3$), 21GAA(Glu, -$x_3$), 29AAG(Lys, -$x_7$). This set coincides with the set of the male links of the α-chain with the exception of the link № 29, which is next to last in the β-chain.

In contrast to the α-chain, where all 4 male triplets and acids correspond to the YY-coordinate "-$x_3$", the coordinate "-$x_3$" of the last male link in the β-chain is replaced by the coordinate "-$x_7$". It breaks one of the male pairs: the triplet AAG, which encodes Lys, exists here instead of the triplet CAG, which differs by the first letter only and which encodes the acid Gln. The second male pair of the links №№ 13 and 21 submits to the phenomenon of doubling because the triplet GAA encodes the acid Glu in both these links.

Whether this phenomenon of doubling is connected with the 3D-construction of insulin (and of those proteins, which possess the same phenomenon) by means of any regular metric or vector relations in space dispositions of pairs of such links? How much widely and precisely the phenomenon of doubling in male and female sub-sets is carried out for different proteins? Many such questions arise in the result of the analysis of objects of molecular genetics from the viewpoint of the genetic octave Yin-Yang-algebra. They should be investigated in future.

## 4 About the origin of the genetic code

Molecular genetics has put forward the question about the origin of the genetic code. The main three version of its origin are considered usually in connection with a stochastic process of molecular evolution (see, for example, [Ratner, 2002, p. 199-202]): 1) the properties of the genetic code could be preset (be predetermined) by physical and chemical characteristics of components and conditions; 2) or they could be selected as adaptive among alternative variants; 3) or they could be fixed accidentally.

For example, the famous hypothesis of a frozen case by F.Crick supposed the following. That accidental but satisfactory system of coding has been fixed by the first in evolutionary process, which further has been multiplied and has undergone to evolutionary complication and optimization in connection with the task of rapid reproduction.

In the author's opinion, the described results of algebraic investigation of the genetic code testify in favor of the fourth version. This fourth version says that the properties of the genetic code were predefined not by physical and chemical, but algebraic characteristics, which are connected with noise immunity coding. From the viewpoint of this fourth version, the biological world (and perhaps not only biological) is encoded in its genetic fundamentals by means of algebraic laws. This is connected with algebraic theory of coding or, perhaps, with the modern algebraic interpretation of the famous idea by Pythagoras: "All things are multi-dimensional numbers".

In what degree the natural process of creation of algebraic bases of the genetic code was stochastic? Whether it is possible, that these bases of genetic coding have been formed not by stochastic way, but they have been, for example, a direct consequence of certain non-stochastic principles (of informational principles?) of the organization of the universe? These questions seem interesting. They are a subject to studying in the future.

One should take into attention the following additional circumstance. The matrix forms of presentation of elements of the genetic octave Yin-Yang-algebra are connected with Hadamard matrices by means of the simple U-algorithm [Petoukhov, arXiv:0802.3366, arXiv:0803.0888, arXiv:0803.3330]. Hadamard matrices play a significant role in the theory of quantum computers and of quantum mechanics, in particular. In this reason such connection seems to be important for possible understanding the systems of the genetic code as quantum mechanical or quantum computer systems. Revealed algebraic properties of the genetic code gives the opportunity to put forward the interesting problem of algebraization of bioinformatics on the basis of the algebras of the genetic code.

The main task of the mathematical natural sciences is the creation of such mathematical models of the natural systems, which can describe these systems in adequate manner. One can see that the algebraic model $YY_8$, which is proposed by the author for the genetic code, fits this task.

It seems that our algebraic-genetic investigations can be useful in some practical tasks as well, for example, in the task of selection of the appropriate sex-partner for an individual by means of personal analysis of molecular-genetic structures of different persons. Similar tasks are not the fantasy but they exist already on world market of genetic services (see for example the website https://www.23andme.com/).

**Acknowledgments**: Described researches were made by the author in the frame of a long-term cooperation between Russian and Hungarian Academies of Sciences and in the frame of programs of "International Society of Symmetry in Bioinformatics" (USA, http://polaris.nova.edu/MST/ISSB) and of "International Symmetry Association" (Hungary, http://symmetry.hu/). The author is grateful to Frolov K.V., Darvas G., Kappraff J., Ne'eman Y., He M., Bakhtiarov K.I., Kassandrov V.V., Smolianinov V.V., Vladimirov Y.S. for their support.